# Enabling Competitive Performance of Medical Imaging with Diffusion Model-generated Images without Privacy Leakage

Yongyi Shi[1], PhD, Wenjun Xia[1], PhD, Chuang Niu[1], PhD, Christopher Wiedeman[1], MD-PhD, Ge Wang[1,2,*], PhD

[1]Biomedical Imaging Center, Rensselaer Polytechnic Institute, Troy, NY, USA 12180

[2]Department of Radiology, Wake Forest University, Winston-Salem, SC, USA 27109

**Deep learning methods have impacted almost every research field, demonstrating notable successes in medical imaging tasks such as denoising and super-resolution. However, the prerequisite for deep learning is data at scale, but data sharing is expensive yet at risk of privacy leakage. As cutting-edge AI generative models, diffusion models have now become dominant because of their rigorous foundation and unprecedented outcomes. Here we propose a latent diffusion approach for data synthesis without compromising patient privacy. In our exemplary case studies, we develop a latent diffusion model to generate medical CT, MRI and PET images using publicly available datasets. We demonstrate that state-of-the-art deep learning-based denoising/super-resolution networks can be trained on our synthetic data to achieve image quality equivalent to what the same network can achieve after being trained on the original data (the *p* values well exceeding the threshold of 0.05). In our advanced diffusion model, we specifically embed a safeguard mechanism to protect patient privacy effectively and efficiently. Consequently, every synthetic image is guaranteed to be different by a pre-specified threshold from the closest counterpart in the original patient dataset. Our approach allows privacy-proof public sharing of diverse big datasets for development of deep models, potentially enabling federated learning at the level of input data instead of local network weights.**

***Corresponding author:** wangg6@rpi.edu.

**Author contributions**

G. W. initiated the project. G. W. and Y. S. designed the experiments. Y. S., W. X. and C. N. performed deep learning research. Y. S. and G. W. wrote the paper, and W. X. and C. W. edited the paper.

**Competing interests**

The authors have no competing interests to declare.

AI methods especially deep learning technologies demonstrate amazing abilities[1] by harnessing big datasets (such as ImageNet[2] for natural image processing) and employing large models (such as ChatGPT[3] and diffusion models), which promises to revolutionize healthcare[4]. However, the imperative to preserve patient privacy[5] has been a challenge for data sharing and hindered the development of big data-empowered AI models in the medical field,[6] such as for healthcare metaverse[7]. Governed by privacy legislation[8], including the United States Health Insurance Portability and Accountability Act (HIPAA)[9] and the European General Data Protection Regulation (GDPR)[10], patient privacy is legally protected, posing a major challenge to medical data sharing and deep learning-based medical imaging. Adherence to ethical, moral, legal and scientific guidelines is crucial to advancing science while respecting privacy[11], complicating the collection of medical data, which is both intricate and costly. Consequently, public medical datasets are limited in size and diversity, typically containing only a few thousand cases at most and being relevant only to a limited number of clinical problems.

To address the huge gap between the imperative for data sharing and the concern for privacy preservation, researchers developed software frameworks for federated learning[12, 13]. Federated learning involves training machine learning models in a decentralised manner without sharing raw data and images. However, recent studies showed that without additional privacy-enhancing techniques, federated learning can be reverse-engineered to reconstruct high-fidelity images from shared gradient weights of the neural networks[14, 15]. To truly ensure patient privacy, federated learning requires additional measures for privacy protection. One technique is differential privacy[16, 17], which adds noise to the trained model weights, but this strategy necessitates careful control between privacy protection and data utility. An ideal approach to overcoming this limitation is to synthesize data that adheres to the same distribution without any quality loss (noising neither data nor parameters) to not only protect privacy but also augment the dataset.

Deep learning-based synthesis methods are highly desirable to learn the distribution of original data in a target domain and generate artificial yet realistic yet diverse samples, while maintaining patient privacy.[18] Among these methods, generative adversarial networks (GANs) are widely used for medical image synthesis,[19, 20] facilitating data sharing,[21] addressing legal or ethical concerns,[22] reducing data requirements through modality translation,[23] and improving deep learning performance.[24] However, GANs have well-known weaknesses, such as mode collapse and unstable training.[25] As a result, generating high-quality medical images remains an open challenge for numerous medical imaging tasks.

Recently, the denoising diffusion probabilistic model (DDPM)[26] including the latent diffusion model (LDM)[27] demonstrated superior performance compared to GANs in synthesis of natural images.[28] The diffusion model transforms input data into noisy data by gradually adding Gaussian noise and then iteratively regenerates the data distribution from pure noise, effectively recovering subtle image details and structures of various types. Diffusion models were then applied to biomedical image synthesis in various areas, including brain magnetic resonance imaging (MRI),[29, 30] histopathology,[31] chest radiographs,[32, 33] and eye funds.[33] Although the diffusion models exhibit great potential for medical image synthesis, privacy guarantees[34] must be developed for their use in privacy-sensitive applications, such as those in which HIPAA applies. Up to now, the feasibility of using the diffusion model for privacy protection in medical image synthesis has not been explored, which is important for downstream medical imaging tasks, such as low-dose computed tomography (LDCT) denoising, MRI super-resolution and positron emission tomography (PET) super-resolution, as well as the development of healthcare metaverse.

Over the past few decades, downstream medical imaging tasks have been a vibrant area of research, spurred by the concerns over patient exposure to ionizing radiation of CT and the needs for enhanced image resolution/quality in MRI and PET. Recently, deep learning techniques demonstrated state-of-the-art performance in denoising,[35, 36] and super-resolution.[37] These techniques typically employ neural networks to learn the relationship between low-quality images and their paired high-quality counterparts. Deep learning-based denoising methods for LDCT can be grouped into projection domain pre-processing,[38, 39] analytic unrolling,[40, 41] dual domain,[42, 43] deep prior-based iterative reconstruction,[44,45] model-based unrolling,[46,47] and image post-processing methods.[48,49,50] On the other hand, deep learning-based super-resolution methods in MRI and PET primarily concentrate on image post-processing techniques.[51,52,53,54] For the development of these downstream medical imaging tasks, a sufficiently large dataset is imperative, and protecting patient privacy is crucial.

In this paper, our overall goal is to synthesize novel realistic data under the original distribution while protecting patient privacy. Achieving this goal will allow that equivalent patient data can be shared directly, which means that the federated learning framework can be upgraded from sharing network parameters to accessing privacy-protected synthetic equivalent counterparts of all available datasets. For this purpose, we adapt the LDM to synthesize realistic medical images from original clinical medical data. To demonstrate the potential of such synthetic data in the downstream medical imaging tasks, we train several representative deep learning models using the public medical dataset or the synthetic counterpart. The results indicate that these models exhibit similar performance after the networks are trained on either original or synthetic datasets, which supports the feasibility of using synthetic data for the downstream medical imaging tasks. It is underlined that when we synthesize data, we enforce the privacy protection inside the LDM by efficiently blocking those synthetic images that closely resemble the original images, providing an effective defense against membership inference attacks.

Our main contributions are as follows:

1) Development of a LDM to generate high-quality medical images that obey the same distribution of original data, with a secured privacy preservability using an efficient latent-space safeguard,

2) Evaluation of several state-of-the-art deep learning-based models trained on the synthetic data, showing equivalent performance in the downstream medical imaging tasks including LDCT denoising and MRI, PET super-resolution.

## Results

### Synthetic Data Inspection

We utilized the LDM to generate synthetic datasets from the Mayo low-dose CT challenge dataset,[55] IXI brain MRI dataset,[56] and brain PET image analysis and disease prediction challenge dataset,[57] achieving Fréchet Inception Distance (FID) scores of 31, 59 and 66, respectively. Subsequently, we assessed the feasibility of training neural networks for LDCT reconstruction, MRI, and PET super-resolution on the respective synthetic datasets. The capability to generate high-quality, novel, and diverse images is crucial for subsequent downstream tasks. As depicted in Fig. 1c, we selected six representative images from the involved imaging modalities, including chest CT, brain MRI, and brain PET, from both the original and synthetic datasets. In each synthetic image, we identified its nearest-matched image in the original dataset for comparison. The synthetic images exhibit global consistency with their corresponding original images and local structural differences. For example, the lung textural features in the original and synthetic CT images are both realistic and yet not identical. The synthetic MRI and PET images exhibit a higher degree of consistency with their respective original images. Indeed, there are anatomically reasonable variations. These variations enhance the diversity of the synthetic dataset, while the global consistency suggests high image quality.

### Privacy Protection

For the sake of patient privacy, we specifically developed a privacy-enhanced LDM (PE-LDM) method to prevent the generation of synthetic images that disclose patient privacy. Taking LDCT as an initial example, we searched through the original images to find one that closely resembles a synthetic image according to root-mean-square error (RMSE). In Fig. 2b, we show the histogram of RMSE between synthetic images and those resembling original images in the cases of LDM and PE-LDM respectively. Our results demonstrate that the minimal RMSE of PE-LDM is larger than that of LDM, indicating that synthetic images generated by PE-LDM are significantly dissimilar to the original images. Fig. 2c presents three samples from the LDM, where the original and synthetic datasets have the minimal RMSE but still differ in detail. The left pair of the images are more similar than the other two pairs. While the middle pair of images are substantially different, the right pair of images are completely different. In contrast, the distribution of images obtained using PE-LDM shows features evidently different yet anatomically meaningful, demonstrating the feasibility of defending against membership inference attacks with principled privacy protection. In Fig. 2d, we show the results of training RED-CNN[48] on synthetic datasets generated by LDM and PE-LDM respectively. Our results demonstrate that PE-LDM offers much-enhanced privacy protection with little compromise in denoising performance; for example, RED-CNN performed equally well with either of the synthetic datasets, as evidenced by a *p* value of 0.314.

**Visual Inspection of LDCT Denoising and MRI/PET Super-Resolution**

To illustrate the denoising effect of the representative networks for LDCT reconstruction, we selected slices from the chest and abdomen regions, as depicted in Figs. 3a and 3b respectively, where the magnified regions-of-interest (ROIs) show details in the lungs and abdomen as marked by the rectangles. The filtered back-projection (FBP) LDCT images in Figs. 3a and 3b exhibit severe noise, with a higher noise level in Fig. 3b due to the stronger attenuation of the abdomen than the lungs. This difference is more pronounced in the magnified ROIs: while the lung texture is affected by noise but still identifiable, the details in the abdomen are difficult to discern. Unlike natural images, CT images have non-uniform noise levels in different regions, posing a unique challenge. It is observed that all evaluated networks demonstrated their denoising capabilities to different degrees. The streak artifacts in the chest images were effectively suppressed, and the texture details in the lungs were well preserved. In contrast, the image quality in the abdominal images significantly improved, but the details could not be recovered well due to the severe noise, as shown in the corresponding magnified views. These results highlight that existing methods need improvement for LDCT reconstruction.

Although all the LDCT reconstruction models significantly improve image quality, particularly with respect to textural details in the lung regions, small differences can still be observed between them. For example, the RED-CNN method suppresses the noise and preserves texture details well, but over-smooths small structures. In contrast, the SUNet[59] method enhances these small structures but amplifies some artifacts in the lungs. When comparing these methods trained on the original and synthetic datasets respectively, the denoising results do not show any significant difference. In Fig. 3b, all the methods lose some fine details due to the elevated noise level. The RED-CNN method over-smooths fine details, while the SUNet method distorts some details. When comparing these methods trained on the original and synthetic datasets, the denoising performance is again essentially identical. Readers can find that the results obtained using the same network trained on either the original or synthetic data are virtually the same without any noticeable difference. This visual inspection demonstrates that training the network on original and synthetic datasets makes little difference.

To showcase the super-resolution performance in the MRI and PET scenarios, we conducted a comparative analysis of various super-resolution networks using brain images, as depicted in Figs. 3c and 3d. To highlight specific details, we magnified the ROIs indicated by the rectangles. All the selected methods, including SRCNN[60], SWIN[61], and DDPM[53], exhibited a consistent performance trend in both MRI and PET 4× super-resolution tasks. Notably, low-resolution images in Figs. 3c and 3d introduced jagged artifacts. The SRCNN method mitigated these artifacts, although the overall result remained somewhat blurry. The SWIN method enhanced performance by introducing sharper edges, yet some fine details remained unrecovered. In visual evaluation, the DDPM method generated sharp edges, closely approximating the ground truth; however, it produced tiny details that differed from the ground truth. While different methods displayed varying performance on MRI and PET super-resolution tasks, visual inspections of these results yielded no significant differences from training the models on either original or synthetic data.

**Statistical Analysis**

We further conducted a systematic quantitative analysis by calculating the peak signal-to-noise ratio (PSNR) and structure similarity (SSIM) values for each image in the whole test dataset and compared the averages in Table I. The results demonstrate excellent alignment with our visual inspection. Although all the denoising methods show significant improvement relative to the FBP LDCT reconstruction, different methods demonstrate varied quantitative metrics. FISTA-Net[47] achieved relatively higher PSNR and SSIM by incorporating sinogram information into the deep denoising process. Using an advanced transformer architecture, The SUNet method produced the best metrics. Most importantly, the denoising results using methods trained on either the original or synthetic data are quantitatively equivalent, as indicated by all *p* value exceeding 0.05. This statistical summary establishes the feasibility of using a synthetic dataset for LDCT training.

The results of super-resolution also aligned well with our visual observations. Although all super-resolution methods demonstrated significant improvement over the low-resolution baseline, they exhibited diverse quantitative metrics. The SWIN method achieved higher PSNR and SSIM than SRCNN. Despite DDPM generated visually appealing results, its quantitative metrics were inferior to SRCNN and SWIN due to discrepancies in relevant details. Most importantly, super-resolution methods trained on either original or synthetic data showed quantitatively very similar results, with all *p* value exceeding 0.05. This statistical summary affirms the feasibility of using a synthetic dataset for MRI and PET super-resolution training.

**Discussions**

Our above-described results can be summarized as the following two key points. First, theoretically speaking, the statistical distribution of original data is the same as that of synthetic data produced by our adapted diffusion model with a dedicated safeguard in the latent space which can block privacy leaking effectively and efficiently. Second, practically speaking, the use of either original or synthetic medical images in training deep networks for LDCT denoising or MRI, PET super-resolution does not produce any significant difference in performance.

There are at least two ways to share big data using our approach. The first is to simply share privacy-proof synthetic data. By the design of our adapted diffusion model, all the synthetic data samples are novel and realistic under the same original data distribution. The second way is to share an executable code of a locally trained diffusion model. This approach avoids the need to transfer synthetic data. A user who receives the trained model can generate as many data samples as needed. A caveat is that if the user has the code, he/she may disable the dictionary extracted from original training data and somehow infer partial or full information on some original samples with a small probability but even in that case the user cannot be sure which synthetic image really reflects a patient image. Also, a secure computing technique can be applied to ensure that the executable code cannot be reverse engineered.

After the release of the Mayo clinic dataset, a large number of LDCT denoising networks were developed, and achieved favorable results on this benchmark dataset. However, unlike natural images, different human body regions have different noise levels due to the heterogeneous attenuation background. For instance, the chest region usually has a lower noise level than the abdomen region. This difference in noise characteristics is even more noticeable in ultra-LDCT scans. In this study, we have shown that existing denoising networks produce promising results in the chest region but may lose fine details in the abdomen region given a worsened signal-to-noise ratio. Hence, there still exists a need to improve image quality and minimize X-ray radiation dose, in order to practice the as low as reasonably achievable (ALARA) principle. A promising direction is to leverage large visual and language models and big datasets to advance the state of the art of deep LDCT denoising in multiple tasks simultaneously. As a follow-up project, we plan to generate more comprehensive datasets for LDCT, MRI, and PET for denoising and super-resolution imaging. Our goal is to train an advanced large model to achieve superior and universally applicable denoising and super-resolution performance. Furthermore, we plan to share the privacy-proof data synthesis codes and protocols publicly.

In conclusion, we have developed a novel approach by embedding a privacy safeguard algorithm in the diffusion model so that image synthesis, data augmentation, and privacy protection can be simultaneously achieved. Specifically, we have combined the variational autoencoder and the diffusion model with a dedicated privacy safeguard to synthesize data under the same original data distribution in the context of deep learning LDCT denoising and MRI, PET super-resolution. Our approach has demonstrated encouraging results and can enable federated learning at the synthetic data level, which is more fundamental than feature-level federated learning and ideal for development of healthcare metaverse.

**Methods**

**Variational Autoencoder**

To reduce the computational cost for training our diffusion model, The variational autoencoder (VAE) is employed to compress the images into a latent space. Fig. 1a shows the VAE network architecture. Training the VAE involves two steps: encoding and decoding. In the encoding process, a distribution of real medical images $x$ is mapped to a posterior distribution $q(z|x)$ that is approximated as a normal distribution of mean µ and standard deviation σ where the latent variable $z = \mu + \sigma\varepsilon$, $\varepsilon \sim \mathcal{N}(0,1)$. The encoder can be parametrized by a neural network as $q_\varphi(z|x)$. In the decoding process, VAE generates a new medical image $\hat{x}$ from a latent variable $z$ though the generative model $p(x|z)$. The decoder can be parametrized as another neural network $p_\kappa(x|z)$. The goal of the VAE training is to reconstruct the image $\hat{x}$ close to the original image $x$.

The VAE loss consists of two terms. The first term is the negative Kullback-Leibler (KL) divergence between $q_\varphi(z|x)$ and $p(z)$, where $p(z)$ is a Gaussian distribution. The second term corresponds to the expected log-likelihood of the observation $x$:

$$\mathcal{L}_{VAE} = D_{KL}\big(q_\varphi(z|x)\|p(z)\big) - \mathbb{E}_{q_\varphi(z|x)}[\log p_\kappa(x|z)]. \qquad (1)$$

Once the VAE is trained, the encoder $q_\varphi(z|x)$ encodes a medical image x into a latent representation z, and the decoder $p_\kappa(x|z)$ reconstructs an image $\hat{x}$ from the latent $z$.

**Latent Diffusion Model**

After medical images are compressed using our VAE, the DDPM can be employed to generate a synthetic dataset. The architecture of the LDM is shown in Fig. 1b containing two sequentially trained models: VAE and DDPM. In the first training phase, the VAE is trained to encode the image space into a latent space. During this VAE training phase, the latent space is directly decoded back into the image space. In the second training phase, the DDPM is trained in the latent space defined by the pre-trained VAE. The weights of the VAE are frozen in the DDPM training phase. The compressed data $z$ is denoted as $z_0 = z$ for the DDPM training. The forward process of DDPM gradually adds Gaussian noise upon a compressed sample $z_0 \sim q(z_0)$ over the course of $T$ timesteps according to a variance schedule $\beta_1, \cdots, \beta_T$. The latent variables $z_1, \cdots, z_T$ have the same dimensionality as the starting point $z_0 \sim q(z_0)$. By the nature of the Gaussian distribution, sampling $z_t$ at an arbitrary timestep $t$ can be written as $z_t = \sqrt{\bar{\alpha}_t} z_0 + \sqrt{1 - \bar{\alpha}_t} \epsilon$, $\epsilon \sim \mathcal{N}(0,1)$, where $\alpha_t = 1 - \beta_t$ and $\bar{\alpha}_t = \prod_{i=1}^{t} \alpha_i$. After the forward process, $z_T$ follows a standard normal distribution when $T$ is large enough. Thus, if we know the joint distribution $q(z_{t-1}|z_t)$, we can easily use the reverse process to get a sample from $q(z_0)$ and $z_T \sim \mathcal{N}(0,1)$. However, $q(z_{t-1}|z_t)$ depends on the entire data distribution, making it difficult to calculate. Hence, a neural network was designed to learn a latent data distribution by gradually denoising a Gaussian random variable, which is to learn the reverse process of a fixed Markov chain with length $T$. Specifically, we use a neural network model $\epsilon_\theta(z_t, t)$ to predict the noise $\epsilon$, which has been shown to work by Ho et al.[26]. Consequently, the corresponding objective can be written as

$$\mathcal{L}_{LDM} = E_{z_0, \epsilon \sim \mathcal{N}(0,1), t}[\| \epsilon - \epsilon_\theta(z_t, t) \|_2^2] \qquad (4)$$

with t is uniformly sampled from $\{1, \cdots, T\}$. In this study, the latent space is diffused into Gaussian noise using $t = 1,000$ steps. An UNet model $\epsilon_\theta(z_t, t)$ is trained to predict the noise $\epsilon$ in the latent space. Samples are then generated with a denoising diffusion implicit model (DDIM)[58] and $t = 150$ steps, with the decoder of the VAE in a single pass.

**Privacy Protection**

We utilize the LDM to generate a synthetic dataset for privacy protection, which offers a fundamental advantage over federated learning; there is no need to share original data nor the weights trained using the original data. However, LDM has a small chance to memorize individual training examples[34] (known as the membership inference), thereby still potentially compromising privacy in a small number of instances. To address this concern, we have developed a PE-LDM method that directly eliminates synthetic images that closely resemble original images, serving as a defense against membership inference attacks.

To achieve this goal, we use VGG to convert the images in the original dataset into vectors of length 1,000 and record all these vectors in a dictionary. During the inference step, we can efficiently convert the output of the LDM into vectors of this type. By comparing the output vectors with the vectors in the dictionary and setting a threshold based on the L2 distance, we can block the synthetic images whose L2 distance is smaller than the threshold. Fig. 2a illustrates the mechanism for privacy enhancement.

**Downstream Medical Imaging Tasks**

Our approach for training deep denoising and super-resolution networks involves two essential steps. Initially, we utilize a LDM to generate a synthetic dataset. Subsequently, we separately train the denoising/super-resolution networks on both the original and synthetic datasets. Previously reported studies have demonstrated promising results with these deep learning-based methods when applied to the original dataset, as they effectively learn the relationship between low-quality and high-quality images. The primary objective of this study is to assess the feasibility of training these deep learning-based methods on the synthetic dataset. To achieve this, we compare the denoising/super-resolution performance of the models trained on the synthetic dataset with that of the same models trained on the original dataset. The selected representative LDCT denoising methods include RED-CNN,[48] FISTA-Net,[47] SUNet.[59] Additionally, the selected representative MRI and PET super-resolution methods, including SRCNN,[60] SWIN,[61] and DDPM[53], focus on the image domain. All denoising results obtained with these methods are evaluated through both visual inspection and quantitative assessments. The image quality metrics employed encompass PSNR and SSIM. The statistical analysis employed the independent samples t-test to assess PSNR. A significance threshold of $p < 0.05$ (two-tailed) was applied for all tests. All calculations were performed with a standard personal computer using SPSS software for Windows, version 29.0.2.0.

**Datasets**

The Mayo LDCT challenge dataset, IXI brain MRI dataset and brain PET image analysis and disease prediction challenge dataset were used to individually train our proposed LDM and validate the clinical performance of the LDCT denoising and MRI, PET super-resolution networks, trained on either the original or synthetic data. In the VAE training phase, the high-quality images from original dataset were used to train the networks. Then, these high-quality images were compressed into a latent space to train the DDPM model. Using the trained LDMs, we randomly generated 2,377 CT, 7000 MRI, and 7000 PET synthetic images, each with dimensions of 512×512, 256×256, 256×256. Simultaneously, we selected 2,377 CT, 7000 MRI, and 7000 PET images from their respective original dataset, with 211 CT, 1000 MRI, and 1000 PET images forming the test dataset. For both the original and synthetic datasets, the corresponding LDCT images were produced by adding Poisson noise into the sinograms simulated from the normal-dose CT images. The noise level can be controlled by the number of photons in the reference air scan, which was uniformly set to $10^4$ photons. The FBP algorithm was employed for benchmark image reconstruction. Additionally, the corresponding low-resolution MRI and PET images were produced by 4× down-sampling.

**Network Training**

To compress 512×512 or 256×256 images into a latent space of 64×64 or 32×32 dimensions, we first trained the VAE with 4 channels. Following the VAE training, we froze its weights and trained the DDPM using an UNet architecture for denoising. Both VAE and DDPM were trained at least 100 epochs, and the training process was stopped when the loss did not decrease by 1% relative to the average loss for the previous 30 epochs. We used a fixed batch size of 4.

We used the trained LDM to generate a synthetic dataset, and then evaluated our selected LDCT denoising and MRI, PET super-resolution networks on the original and/or synthetic datasets. In reference to the corresponding original references, we did our best to optimize the parameters of all the methods and maintained the same parameters for each algorithm in the cases of the original and/or synthetic datasets.

## Data availability

The dataset for the NIH-AAPM-Mayo clinic low-dose CT grand challenge is publicly accessible via the following link: http://www.aapm.org/GrandChallenge/LowDoseCT/. The IXI brain MRI dataset is publicly available on https://brain-development.org/ixi-dataset/. The dataset for the brain PET image analysis and disease prediction challenge dataset is publicly available on https://challenge.xfyun.cn/topic/info?type=PET.

## Code availability

The source code is accessible to the public on https://github.com/shiyongyi/LDM_privacy.

## References

1. J. Wei, Y. Tay, R. Bommasani, C. Raffel, B. Zoph, S. Borgeaud, et al., "Emergent abilities of large language models," arXiv preprint arXiv:2206.07682. 2022.

2. J. Deng, W. Dong, R. Socher, L. Li, K. Li, Fei-Fei, L., "ImageNet: A large-scale hierarchical image database," IEEE conference on computer vision and pattern recognition, pp. 248-255, 2009.

3. OpenAI. Gpt-4 technical report, arXiv preprint arXiv:2303.08774, 2023.

4. M. Karatas, L. Eriskin, M. Deveci, D. Pamucar, H. Garg, "Big Data for Healthcare Industry 4.0: Applications, challenges and future perspectives," Expert Systems with Applications, vol. 200, p. 116912, 2022.

5. G. Kaissis, M. Makowski, D. Rückert, R. Braren, "Secure, privacy-preserving and federated machine learning in medical imaging," Nature Machine Intelligence, vol. 2, no. 6, pp. 305-311, 2020.

6. W. Price, I. Cohen, "Privacy in the age of medical big data," Nature medicine, vol. 25, no. 1, pp. 37-43, 2019.

7. G. Wang, A. Badal, X. Jia, J. Maltz, K. Mueller, K. Myers, et al., "Development of metaverse for intelligent healthcare," Nature Machine Intelligence, vol. 4, pp. 922-929, 2022.

8. H. Jin, Y. Luo, P. Li, J. Mathew, "A review of secure and privacy-preserving medical data sharing," IEEE Access, vol. 7, pp. 61656-61669, 2019.

9. HIPAA. US Department of Health and Human Services https://www.hhs.gov/hipaa/index.html, 2020.

10. GDPR. Intersof Consulting https://gdpr-info.eu, 2016.

11. D. Usynin, A. Ziller, M. Makowski, R. Braren, D. Rueckert, B. Glocker, et al., "Adversarial interference and its mitigations in privacy-preserving collaborative machine learning," Nature Machine Intelligence, vol. 3, no. 9, pp. 749-758, 2021.

12. N. Rieke, J. Hancox, W. Li, F. Milletari, H. Roth, S. Albarqouni, et al., "The future of digital health with federated learning," NPJ digital medicine, vol. 3, no. 1, p. 119, 2020.

13. M. Adnan, S. Kalra, J. Cresswell, G. Taylor, H. Tizhoosh, "Federated learning and differential privacy for medical image analysis," Scientific reports, vol. 12, no. 1, p. 1953, 2022.

14. C. Schwarz, W. Kremers, T. Therneau, R. Sharp, J. Gunter, P. Vemuri, et al., "Identification of anonymous MRI research participants with face-recognition software," New England Journal of Medicine, vol. 381, no. 17, pp. 1684-1686, 2019.

15. J. Geiping, H. Bauermeister, H. Dröge, M. Moeller, "Inverting gradients-how easy is it to break privacy in federated learning?" Advances in Neural Information Processing Systems, vol. 33, pp. 16937-16947, 2020.

16. G. Kaissis, A. Ziller, J. Passerat-Palmbach, T. Ryffel, D. Usynin, A. Trask, et al., "End-to-end privacy preserving deep learning on multi-institutional medical imaging," Nature Machine Intelligence, vol. 3, no. 6, pp. 473-484, 2021.

17. A. Ziller, D. Usynin, R. Braren, M. Makowski, D. Rueckert, G. Kaissis, "Medical imaging deep learning with differential privacy," Scientific Reports, vol. 11, p. 13524, 2021.

18. T. Wang, Y. Lei, Y. Fu, J. F. Wynne, W. J. Curran, T. Liu, X. Yang, "A review on medical imaging synthesis using deep learning and its clinical applications," Journal of Applied Clinical Medical Physics, vol. 22, no. 1, pp. 11-36, 2021.

19. Y. Chen, X. H. Yang, Z. Wei, A. A. Heidari, N. Zheng, Z. Li, et al., "Generative adversarial networks in medical image augmentation: a review," Computers in Biology and Medicine, vol. 144, p. 105382, 2022.

20. C. Gao, B. Killeen, Y. Hu, R. Grupp, R. Taylor, M. Armand, et al., "Synthetic data accelerates the development of generalizable learning-based algorithms for X-ray image analysis," Nature Machine Intelligence, vol. 5, pp. 294-308, 2023.

21. A. DuMont Schütte, J. Hetzel, S. Gatidis, T. Hepp, B. Dietz, S. Bauer, et al., "Overcoming barriers to data sharing with medical image generation: a comprehensive evaluation," NPJ digital medicine, vol. 4, no. 1, p. 141, 2021.

22. T. Han, S. Nebelung, C. Haarburger, N. Horst, S. Reinartz, D. Merhof, et al., "Breaking medical data sharing boundaries by using synthesized radiographs," Science Advances, vol. 6, no. 49, p. eabb7973, 2020.

23. K. Armanious, C. Jiang, M. Fischer, T. Küstner, T. Hepp, K. Nikolaou, et al., "MedGAN: Medical image translation using GANs," Computerized Medical Imaging and Graphics, vol. 79, p. 101684, 2020.

24. M. Frid-Adar, I. Diamant, E. Klang, M. Amitai, J. Goldberger, H. Greenspan, "GAN-based synthetic medical image augmentation for increased CNN performance in liver lesion classification," Neurocomputing, vol. 321, pp. 321-331, 2018.

25. D. Saxena, J. Cao, "Generative adversarial networks (GANs) challenges, solutions, and future directions," ACM Computing Surveys (CSUR), vol. 54, no. 3, pp. 1-42, 2021.

26. J. Ho, A. Jain, P. Abbeel, "Denoising diffusion probabilistic models. Advances in Neural Information Processing Systems," vol. 33, pp. 6840-51, 2020.

27. R. Rombach, A. Blattmann, D. Lorenz, P. Esser, B. Ommer, "High-resolution image synthesis with latent diffusion models," In Proceedings of the IEEE/CVF Conference on Computer Vision and Pattern Recognition, pp. 10684-10695, 2022.

28. P. Dhariwal, A. Nichol, "Diffusion models beat GANs on image synthesis," Advances in Neural Information Processing Systems, vol. 34, pp. 8780-8794, 2021.

29. W. H. Pinaya, P. D. Tudosiu, J. Dafflon, P. F. Da Costa, V. Fernandez, P. Nachev et al., "Brain imaging generation with latent diffusion models," In MICCAI Workshop on Deep Generative Models pp. 117-126, 2022.

30. W. Peng, E. Adeli, Q. Zhao, K. M. Pohl, "Generating Realistic 3D Brain MRIs Using a Conditional Diffusion Probabilistic Model," arXiv preprint arXiv:2212.08034. 2022.

31. P. A. Moghadam, S. Van Dalen, K. C. Martin, J. Lennerz, S. Yip, H. Farahani, "A Morphology Focused Diffusion Probabilistic Model for Synthesis of Histopathology

Images," In Proceedings of the IEEE/CVF Winter Conference on Applications of Computer Vision, 2023.

32. K. Packhäuser, L. Folle, F. Thamm, A. Maier, "Generation of anonymous chest radiographs using latent diffusion models for training thoracic abnormality classification systems," arXiv preprint arXiv:2211.01323. 2022.

33. G. Müller-Franzes, J. M. Niehues, F. Khader, S. T. Arasteh, C. Haarburger, C. Kuhl et al., "Diffusion Probabilistic Models beat GANs on Medical Images," arXiv preprint arXiv:2212.07501. 2022.

34. N. Carlini, J. Hayes, M. Nasr, M. Jagielski, V. Sehwag, F. Tramer, et al., "Extracting training data from diffusion models," arXiv preprint arXiv:2301.13188. 2023.

35. G. Wang, J. C. Ye, B. De Man, "Deep learning for tomographic image reconstruction," Nature Machine Intelligence, vol. 2, no. 12, pp. 737-748. 2020.

36. G. Wang, M. Jacob, X. Mou, Y. Shi, Y. C. Eldar, "Deep Tomographic Image Reconstruction: Yesterday, Today, and Tomorrow—Editorial for the 2nd Special Issue "Machine Learning for Image Reconstruction"," IEEE Transactions on Medical Imaging, vol. 40, no. 11, pp. 2956-2964, 2021.

37. Y. Li, B. Sixou, F. Peyrin, "A review of the deep learning methods for medical images super resolution problems," IRBM, vol. 42, no. 2, pp. 120-133, 2021.

38. Y. J. Ma, Y. Ren, P. Feng, P. He, X. D. Guo, B. Wei, "Sinogram denoising via attention residual dense convolutional neural network for low-dose computed tomography," Nuclear Science and Techniques, vol. 32, no. 41, pp. 1-14, 2021.

39. L. Yang, Z. Li, R. Ge, J. Zhao, H. Si, D. Zhang, "Low-Dose CT Denoising via Sinogram Inner-Structure Transformer," IEEE Transactions on Medical Imaging, early access, 2022.

40. B. Zhu, J. Z. Liu, S. F. Cauley, B. R. Rosen, M. S. Rosen, "Image reconstruction by domain-transform manifold learning," Nature, vol. 555, no. 7697, pp. 487-492, 2018.

41. Y. Li, K. Li, C. Zhang, J. Montoya, G. H. Chen, "Learning to reconstruct computed tomography images directly from sinogram data under a variety of data acquisition conditions," IEEE Transactions on Medical Imaging, vol. 38, no. 10, pp. 2469-2481, 2019.

42. Y. Ge, T. Su, J. Zhu, X. Deng, Q. Zhang, J. Chen, et al., "ADAPTIVE-NET: deep computed tomography reconstruction network with analytical domain transformation knowledge," Quantitative Imaging in Medicine and Surgery, vol. 10, no. 2, pp. 415-427, 2020.

43. D. Hu, J. Liu, T. Lv, Q. Zhao, Y. Zhang, G. Quan et al., "Hybrid-domain neural network processing for sparse-view CT reconstruction," IEEE Transactions on Radiation and Plasma Medical Sciences, vol. 5, no. 1, pp. 88-98, 2020.

44. D. Wu, K. Kim, G. El Fakhri, Q. Li, "Iterative low-dose CT reconstruction with priors trained by artificial neural network," IEEE Transactions on Medical Imaging, vol. 36, no. 12, pp. 2479-2486, 2017.

45. S. Ye, Z. Li, M. T. McCann, Y. Long, S. Ravishankar, "Unified supervised-unsupervised (super) learning for x-ray CT image reconstruction," IEEE Transactions on Medical Imaging, vol. 40, no. 11, pp. 2986-3001, 2021.

46. J. Adler, O. Öktem, "Learned primal-dual reconstruction," IEEE Transactions on Medical Imaging, vol. 37, no. 6, pp. 1322-1332, 2018.

47. J. Xiang, Y. Dong, Y. Yang, "FISTA-net: Learning a fast iterative shrinkage thresholding network for inverse problems in imaging," IEEE Transactions on Medical Imaging, vol. 40, no. 5, pp. 1329-1339, 2021.

48. H. Chen, Y. Zhang, M. K. Kalra, F. Lin, Y. Chen, P. Liao et al., "Low-dose CT with a residual encoder-decoder convolutional neural network," IEEE Transactions on Medical Imaging, vol. 36, no. 12, pp. 2524-35, 2017.

49. K. H. Jin, M. T. McCann, E. Froustey, M. Unser, "Deep convolutional neural network for inverse problems in imaging," IEEE Transactions on Image Processing, vol. 26, no. 9, pp. 4509-4522, 2017.

50. H. Shan, A. Padole, F. Homayounieh, U. Kruger, R. D. Khera, C. Nitiwarangkul, G. Wang, "Competitive performance of a modularized deep neural network compared to commercial algorithms for low-dose CT image reconstruction," Nature Machine Intelligence, vol. 1, no. 6, pp. 269-276, 2019.

51. Q. Lyu, H. Shan, C. Steber, C. Helis, C. Whitlow, M. Chan, et al., "Multi-contrast super-resolution MRI through a progressive network," IEEE Transactions on Medical Imaging, vol. 39, no. 9, pp. 2738-2749, 2020.

52. G. Li, J. Lv, Y. Tian, Q. Dou, C. Wang, C. Xu, et al., "Transformer-empowered multi-scale contextual matching and aggregation for multi-contrast MRI super-resolution," In Proceedings of the IEEE/CVF Conference on Computer Vision and Pattern Recognition, pp. 20636-20645, 2022.

53. Z. Wu, X. Chen, S. Xie, J. Shen, Y. Zeng, "Super-resolution of brain MRI images based on denoising diffusion probabilistic model," Biomedical Signal Processing and Control, vol. 1, no. 85, p. 104901, 2023.

53. J. A. Kennedy, O. Israel, A. Frenkel, R. Bar-Shalom, H. Azhari, "Super-resolution in PET imaging," IEEE Transactions on Medical Imaging, vol. 25, no. 2, pp. 137-147, 2006.

54. T. A. Song, S. R. Chowdhury, F. Yang, J. Dutta, "Super-resolution PET imaging using convolutional neural networks," IEEE Transactions on Computational Imaging, vol. 6, pp. 519-528, 2020.

55. [Online] Available: https://www.aapm.org/grandchallenge/lowdosect/

56. [Online] Available: https://brain-development.org/ixi-dataset/

57. [Online] Available: https://challenge.xfyun.cn/topic/info?type=PET

58. J. Song, C. Meng, S. Ermon, "Denoising diffusion implicit models," arXiv preprint arXiv:2010.02502. 2020.

59. C. Fan, T. Liu, K. Liu, "SUNet: swin transformer UNet for image denoising," In IEEE International Symposium on Circuits and Systems (ISCAS), pp. 2333-2337, 2022.

60. C. Dong, C. C. Loy, K. He, X. Tang, "Learning a deep convolutional network for image super-resolution," In Computer Vision–ECCV 2014, pp. 184-199, 2014.

61. A. Sadikov, J. Wren-Jarvis, X. Pan, L. T. Cai, P. Mukherjee, "Generalized Diffusion MRI Denoising and Super-Resolution using Swin Transformers," arXiv preprint arXiv:2303.05686. 2023.

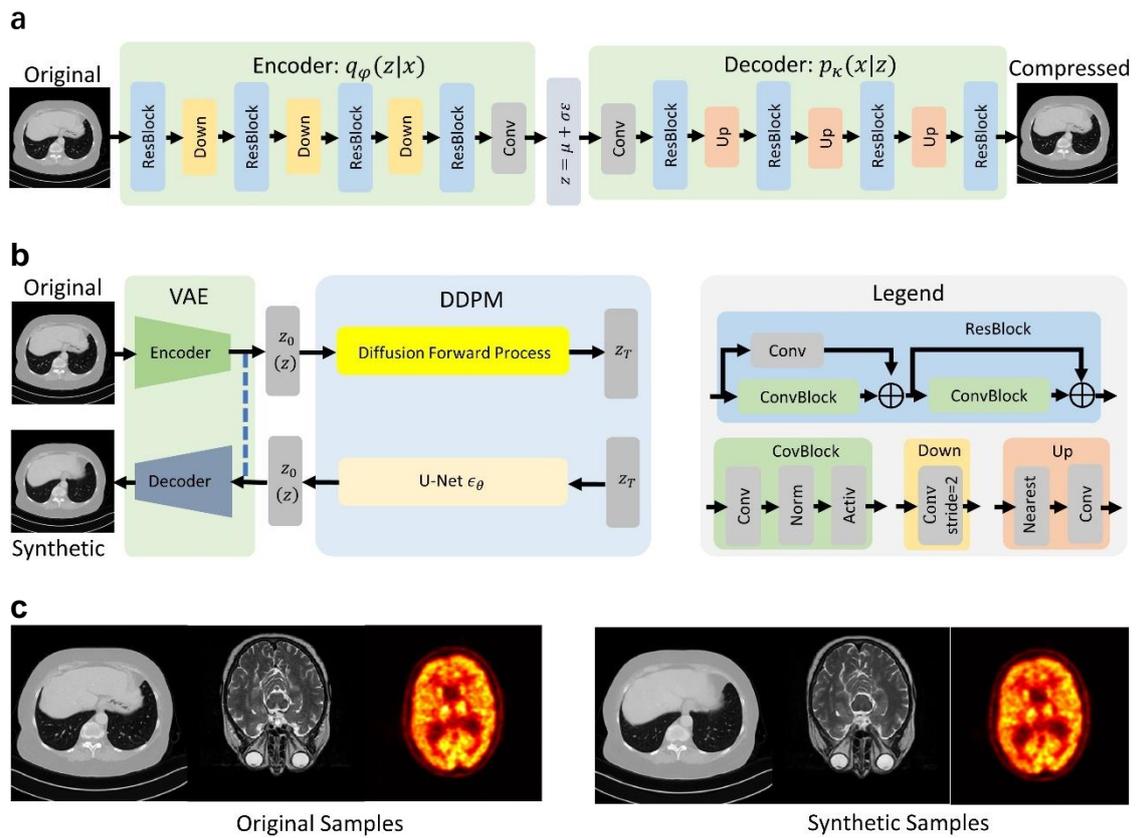

**Figure 1: The architecture of LDM and six sample images. (a)** the VAE architecture, **(b)** the LDM architecture, and **(c)** six sample images from the chest CT, brain MRI and brain PET in the original and synthetic datasets respectively.

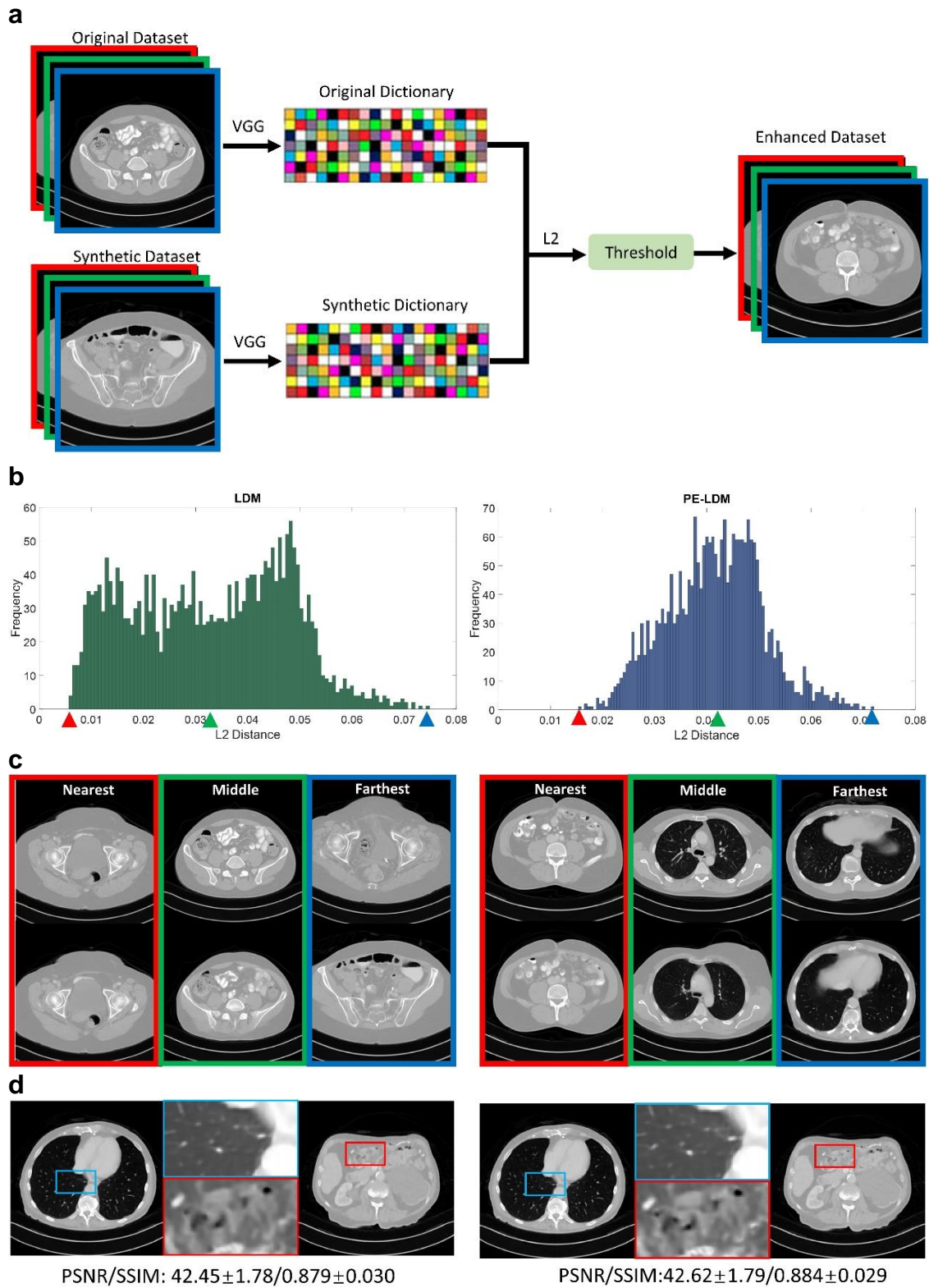

**Figure 2: Privacy analysis for LDM and PE-LDM**. **(a)** schematic for privacy enhancement. **(b)** The histograms for LDM (left) and PE-LDM (right) respectively. **(c)** sample images for visual inspection of privacy protection, and **(d)** the equivalent results denoised using RED-CNN networks independently trained on the datasets synthesized by LDM and PE-LDM respectively, with the PSNR and SSIM values computed on the whole test dataset.

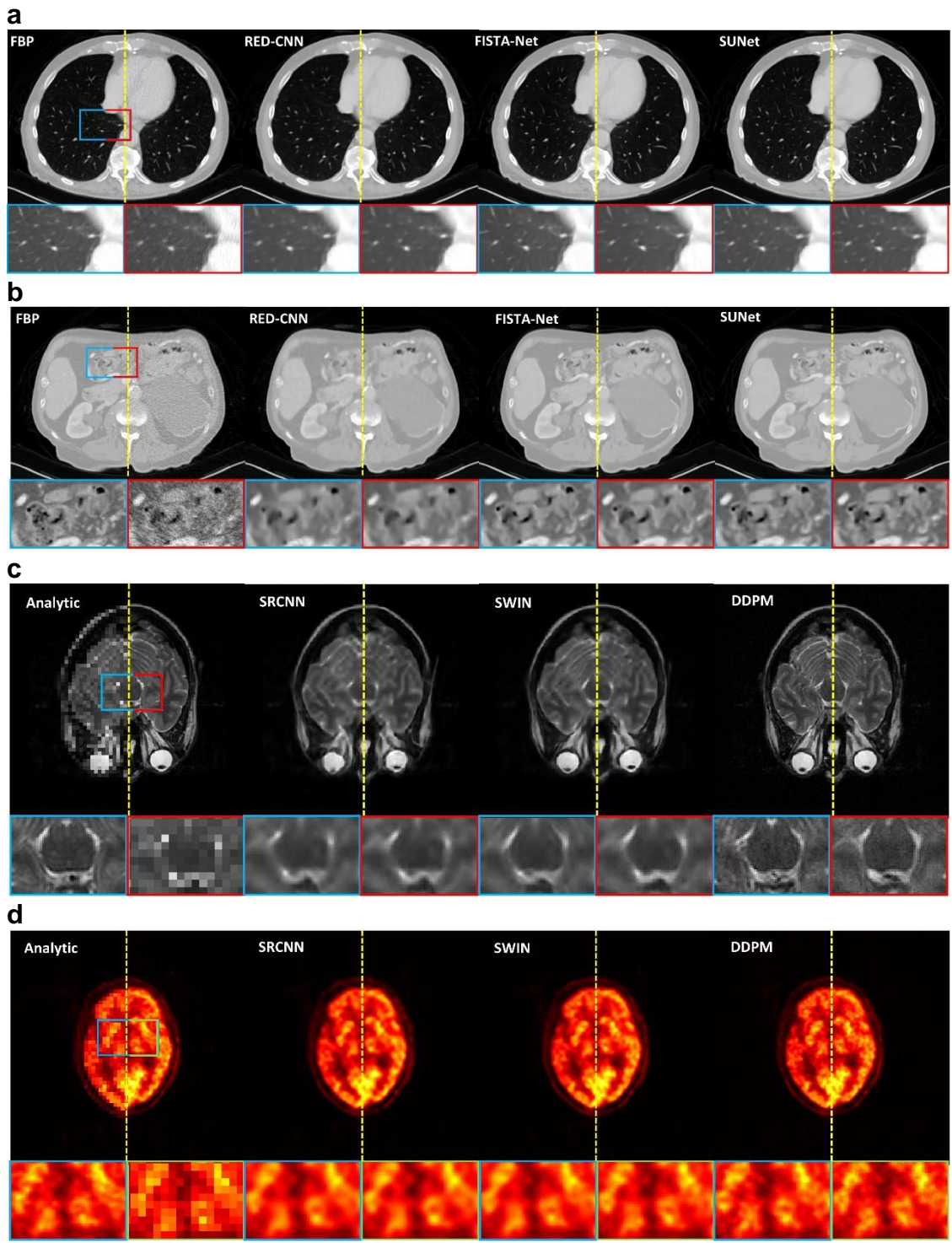

**Figure 3: Image results for low-dose CT denoising and MRI/PET super-resolution. (a)** Chest CT denoising images, each of which is split in half to shown more information – for FBP image, left and right halves are from NDCT and LDCT respectively; for other methods, left and right halves are from original

and synthetic datasets respectively, and the zoom-in views of the ROIs for each image are colour-boxed to show the results denoised based on the original and synthetic data respectively, **(b)** abdomen CT denoising images presented in the same fashion, **(c)** MRI super-resolution images, each of which is split in half to shown more information – for analytic reconstructed image, left and right halves are from high-resolution and low-resolution respectively; for other methods, left and right halves are from original and synthetic datasets respectively, and the zoom-in views of the ROIs for each image are colour-boxed to show the results denoised based on the original and synthetic data respectively, and **(d)** PET super-resolution images presented in the same fashion.

Table I. Measurements of PSNR and SSIM.

| Modality | Methods | PNSR | | SSIM | | p Value |
|---|---|---|---|---|---|---|
| | | Original | Synthetic | Original | Synthetic | |
| CT | LD-FBP | 36.10±2.34 | - | 0.778±0.055 | - | - |
| | RED-CNN | 42.46±1.80 | 42.45±1.78 | 0.879±0.029 | 0.879±0.030 | 0.936 |
| | FISTA-Net | 43.47±2.02 | 43.26±2.14 | 0.896±0.242 | 0.894±0.263 | 0.303 |
| | SUNet | 44.92±1.89 | 44.77±1.84 | 0.922±0.026 | 0.920±0.027 | 0.416 |
| MRI | Analytic | 21.39±1.06 | - | 0.844±0.028 | - | - |
| | SRCNN | 26.83±1.25 | 26.82±1.28 | 0.936±0.014 | 0.938±0.015 | 0.937 |
| | SWIN | 28.17±1.45 | 28.06±1.49 | 0.952±0.014 | 0.952±0.014 | 0.100 |
| | DDPM | 26.19±1.60 | 26.28±1.59 | 0.913±0.015 | 0.912±0.014 | 0.245 |
| PET | Analytic | 26.85±0.83 | - | 0.856±0.273 | - | - |
| | SRCNN | 38.75±3.94 | 38.69±4.12 | 0.972±0.013 | 0.973±0.014 | 0.729 |
| | SWIN | 39.00±3.60 | 38.78±4.17 | 0.976±0.012 | 0.974±0.014 | 0.208 |
| | DDPM | 37.05±3.32 | 36.80±3.28 | 0.942±0.018 | 0.946±0.016 | 0.079 |

Note. – Unless otherwise specified, data are means ± standard deviations.